\documentclass[doublecol,figures]{epl2}
\usepackage{amsmath}
\usepackage{wasysym}
\usepackage{color}

\title{Tension enhancement in branched macromolecules upon adhesion
on a solid substrate}
\shorttitle{Tension enhancement in branched molecules}

\author{
Jaroslaw Paturej\inst{1,2}\thanks{
E-mail: \email{jpaturej@univ.szczecin.pl}} \and
Lukasz Kuban\inst{3} \and
 Andrey Milchev\inst{1,4} \and Thomas A. Vilgis\inst{1}}
\shortauthor{J. Paturej \etal}

\institute{ \inst{1} Max Planck Institute for Polymer Research, 10 Ackermannweg,
55128 Mainz, Germany, EU\\
\inst{2} Institute of Physics, University of Szczecin, Wielkopolska 15,
70451 Szczecin, Poland, EU\\
\inst{3} Institute of Thermal Machinery, Czestochowa University of Technology,
Armii Krajowej 21, 42200 Czestochowa, Poland, EU\\
\inst{4} Institute for Physical Chemistry, Bulgarian Academy of Sciences, 1113
Sofia, Bulgaria, EU}


\abstract{The effect of self-generated tension in the backbone of a bottle-brush
(BB) macromolecule, adsorbed on an attractive surface, is studied by means
of Molecular Dynamics simulations of a coarse-grained bead-spring model in the
good solvent regime. The BB-molecule is modeled as a backbone chain of $L$
beads, connected by breakable bonds and with side chains, tethered pairwise to
each monomer of the backbone. Our investigation is focused on several key
questions that determine the bond scission mechanism and the ensuing degradation
kinetics: how are frequency of bond scission and self-induced tension
distributed along the BB-backbone at different grafting density $\sigma_g$ of
the side chains? How does tension $f$ depend on the length of the side chains
$N$, and on the strength of surface adhesion $\epsilon_s$? We examine the
monomer density distribution profiles across the BB-backbone at different
$\epsilon_s$ and relate it to adsorption-induced morphological changes of the
macromolecule whereby side chains partially desorb while the remaining
chains spread better on the surface. Our simulation data are found to be in
qualitative agreement with experimental results and recent theoretical
predictions. Yet we demonstrate that the interval of parameter values where
these predictions hold is limited in $N$. Thus,  at high values of $\epsilon_s$,
too long side chains mutually block each other and freeze effectively the
bottle-brush molecule.}

\begin{document}

\maketitle

\section{Introduction}

One of the most outstanding challenges in modern material sciences is the design
and synthesis of "smart" macromolecules with stress-activated functions
\cite{caruso,craig}. During the last decade one observes thus a rapidly
escalating interest in the field of novel polymer mechanochemistry which, in
contrast to the traditional (non-selective) one, allows to control bond tension
on molecular length scales \cite{caruso,beyer,liang,huang}. Examples related to
these advances enable, for instance,  rupture of specific chemical
bonds\cite{park,yang}, steering the course of chemical
reactivity\cite{wiita,hicken,craig2,li}, changing color of materials
\cite{davis} or mapping their stress distribution at molecular level
\cite{cho,clark}.

In a series of recent experiments, a strong enhancement of the tension in the
(typically, polymethacrylate with degree of polymerization $L=3600$) backbone of
bottle-brush polymers with side chains  of poly($n-$butyl acrylate) of length
$N=140$, self-induced upon adsorption on a solid surface (mica), was reported
\cite{park,sheiko,lebedeva,park2}.  An experimental method for control and
manipulation of the bond-cleavage in bottle-brush backbones was also proposed
\cite{park2}. Thus, a selectivity of bond breakage can be achieved by tuning the
molecular size of such macromolecules which makes it possible to fabricate the
brush so as to focus tension in the middle of the molecule. The increase of the
bond tension in these macromolecules is induced by the steric repulsion of the
side chains as they tend to maximize the number of contacts with the substrate
in order to gain energy. This tension, which depends on grafting density
$\sigma_g$, on the side chain length $N$, and on the strength of substrate
attraction $\epsilon_s$, effectively lowers the energy barrier for bond
scission. As observed in experiments, self-induced build up of tension
proves sufficient to instantly sever covalent bond in the backbone.

The possibility for breaking strong covalent bonds is also an interesting
problem from the standpoint of fundamental physics.  Amplification of tension in
branched polymers has been considered theoretically  by Panyukov and
collaborators in several recent works \cite{244,panyukovprl,panyukovmacro} by
means of scaling theory and Self Consistent Field techniques. Numerous possible
regimes of brush-molecule behavior in terms of $N,\; \sigma_g$ and
$\epsilon_s$ have been outlined~\cite{244}.  It was argued that polymers
with branched morphology, physically adsorbed on an attractive plane, allow
focusing of the side-chain tension on the backbone whereby at given temperature
$T$ the tension in the backbone becomes proportional to the length of the side
chain, $f\approx f_{\mbox{\tiny S}}N$. Here $f_{\mbox{\tiny S}}$ denotes the
maximum tension in
the side chains, $f_{\mbox{\tiny S}} \approx k_BT/b$, with $k_{\mbox{\tiny B}}$
being the
Boltzmann constant, and $b$ - the Kuhn length (or, the monomer diameter for
absolutely flexible chains). However, a comprehensive understanding of
covalent bond breaking in adsorbed branched polymers still has to be reached.
Many of the detailed theoretical predictions can hardly be measured directly
experimentally. It remains unclear how exactly tension builds up with
growing adhesion to the solid surface and impact exercise side chain length $N$
and grafting density $\sigma_g$ on this phenomenon. Much  insight in this
respect can be provided by computer experiments, yet there have been only few
such studies~\cite{244,milchev} which didn't focus on these questions.

In an earlier molecular dynamics (MD) simulation study of the scission kinetics
it was found that mean lifetime of a bond becomes more than an order of
magnitude shorter when the bottle-brush molecule spreads on an adhesive surface
\cite{milchev}. These simulation results \cite{milchev} indicated also that the
probability distribution for bond scission along the backbone of a bottle-brush
is sensitive to the grafting density (and, therefore, to  the degree of steric
repulsion) of the side chains. The shape of this distribution resembles the
experimentally established one \cite{sheiko} only for weaker repulsion when the
side chains are short and do not mutually block one another.


In this letter we report our MD results  on the first direct measurement of the
adsorption-induced tension $f$ in the backbone of a bottle-brush macromolecule
whereby we examine the influence of varying parameters: grafting density
$\sigma_g$, adsorption strength $\epsilon_s/k_BT$ as well as the length of side
chains $N$ on the resulting  increase of $f$.

\section{The model}

We consider a three-dimensional coarse-grained model of a bottle-brush
macromolecule which consists $L$ monomers in the backbone connected by bonds.
Moreover, two side chains of length $N$ are grafted to every $\sigma_g^{-1}$-th
repeatable unit of the backbone (except for the terminal beads of the backbone
where there are three side chains anchored). In this way a grafting density
$\sigma_g$, which gives the number of side chains {\it pairs} per unit length,
is defined. Thus, the total number of monomers in $M$ the brush molecule is
$M=L+2N[(L-1)\sigma_g+2]$.
\begin{figure}
\twofigures[scale=0.17]{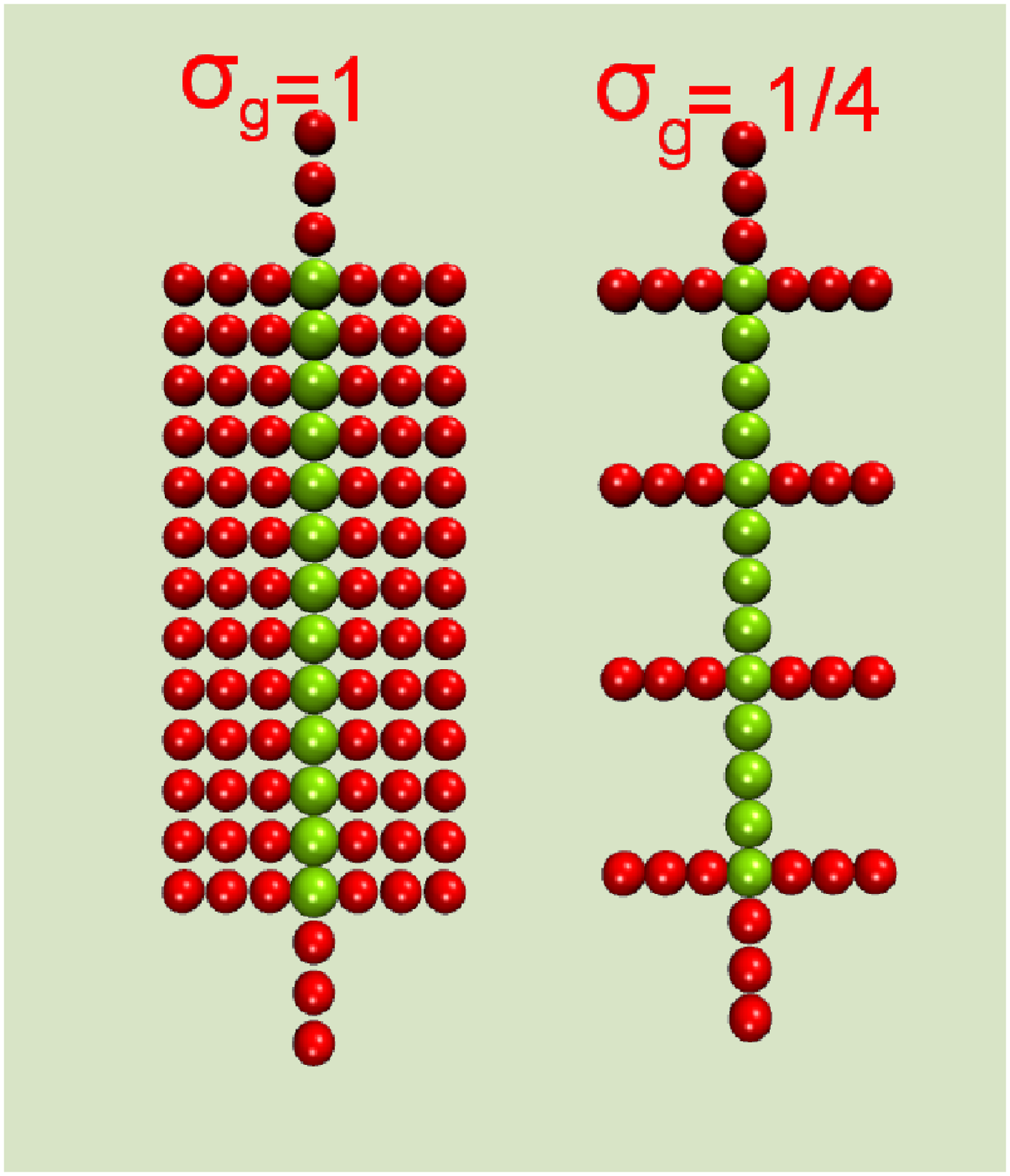}{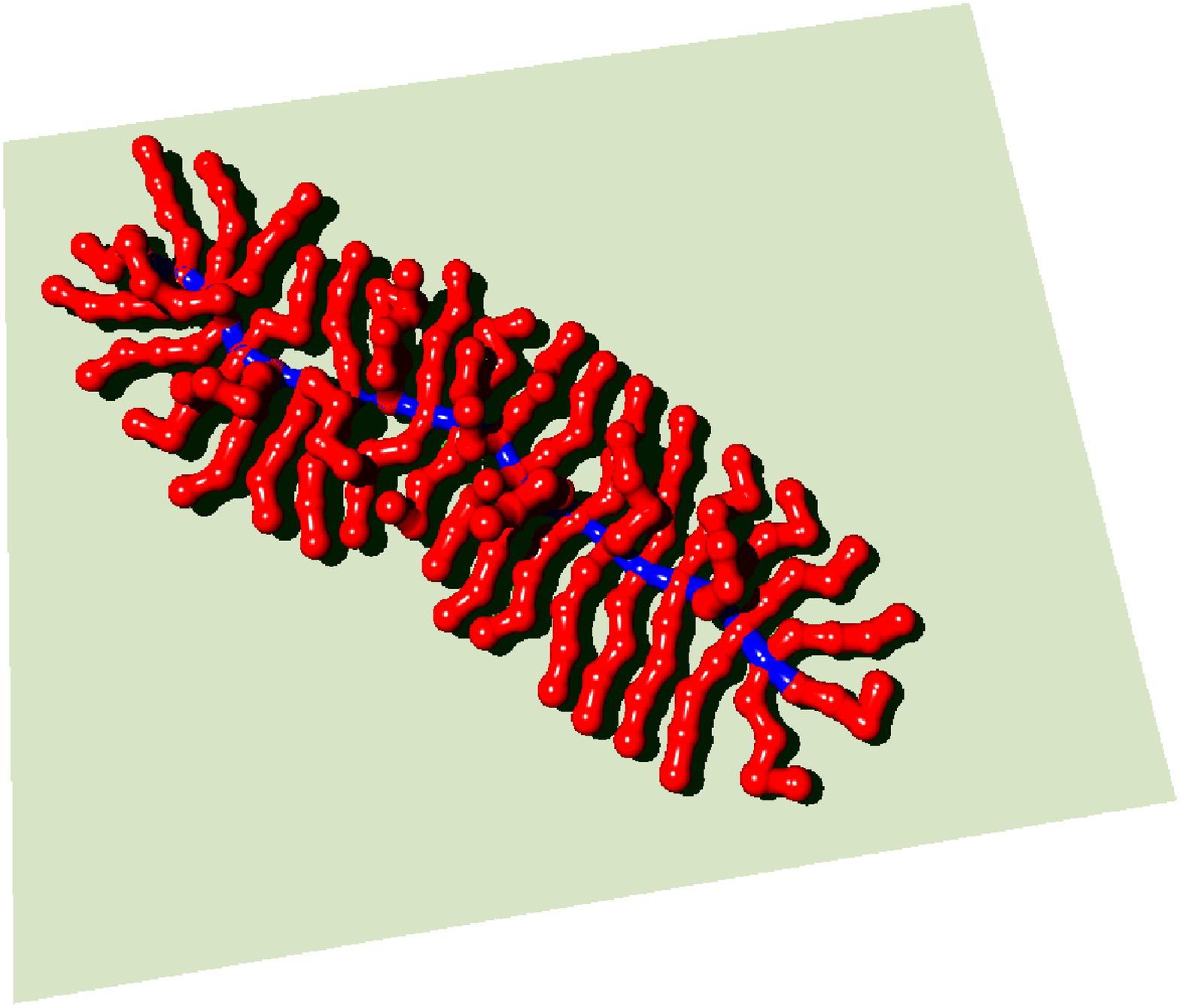}
\caption{ Starting configuration of a bottle-brush molecule (a "centipede") with
$L=13$ (backbone) and $N=3$ (side chain), so that for a grafting density
$\sigma_g=1$, the total number of segments $M=97$, and for $\sigma_g=1/4$, one
has $M=43$. }
\label{flatbb}
\caption{A snapshot of a thermalized "centipede" with $L=20$ backbone monomers
(blue) and $42$ side chains (red) of length $N=4$. Side chains which are too
strongly squeezed by the neighbors (especially when the backbone bends) are seen
to get off the substrate in order to minimize free energy.}
\label{centipede}
\end{figure}
The bonded interaction of each segment in the brush with its neighbors is
described by the frequently used Kremer-Grest potential, $V^{\mbox{\tiny
KG}}(r)=V^{\mbox{\tiny FENE}}+V^{\mbox{\tiny WCA}}$, with the so-called
'finitely-extensible nonlinear elastic' (FENE) potential:
\begin{equation}
 V^{\mbox{\tiny FENE}}(r)=- 0.5kr_0^2\ln{[1-(r/{r_0})^2]},
\label{fene}
\end{equation}
and the Weeks-Chandler-Anderson (WCA) (i.e., the shifted and truncated repulsive
branch of the Lennard-Jones potential) given by:
\begin{equation}
 V^{\mbox{\tiny WCA}}(r) = 4\epsilon\left[
(\sigma/ r)^{12} - (\sigma /r)^6 + 1/4
\right]\theta(2^{1/6}\sigma-r)
\label{wca}
\end{equation}
In Eq.~(\ref{fene}) $k=30$, $r_0=1.5$ whereas in Eq.~(\ref{wca}) one has
$\theta(x)=0$ or 1 for $x<0$ or $x\geq 0$, and $\epsilon=1$, $\sigma=1$. The
potential $V^{\mbox{\tiny KG}}(r)$ has a minimum at bond length $r_{\mbox{\tiny
bond}}\approx 0.96$. Thus, the bonded interaction, $V^{\mbox{\tiny KG}}(r)$,
makes the bonds of the brush in our model unbreakable which is convenient when
measuring bond tension is our main concern. In the course of the study, however,
we have also analyzed the case when {\it backbone} bonds may undergo thermal
scission. The bonded potential for the {\it backbone} monomers is then replaced
by $V^{\mbox{\tiny b}}(r)=V^{\mbox{\tiny M}}+V^{\mbox{\tiny WCA}}$, with a Morse
potential
\begin{equation}
V^{\mbox{\tiny M}}(r) = D\lbrace 1-\exp{[-\alpha(r-b)]} \rbrace^2
\label{morse}
\end{equation}
which was recently used in simulations exploring breakage of polymer chains
\cite{gosh,paturejjcp,paturejepl} and bottle-brushes \cite{milchev}.
In Eq.~(\ref{morse}) $\alpha$ is a constant that determines bond elasticity, we
use here $\alpha=1$. The dissociation energy $D$  of a given bond is measured in
units of $k_BT$, where $k_B$ is the Boltzmann constant and $T$ denotes the
temperature. We use here $D=1$. The monomer diameter in the  case of
"breakable" potential
(\ref{morse}) is $b=2^{1/6}\sigma\approx 1.12\sigma$. All non-bonded
interactions between monomers are taken into account be means of the WCA
potential, Eq.~(\ref{wca}). In consequence the interactions in our model
correspond to good solvent conditions.

\begin{figure}[htb]
\onefigure[scale=0.35]{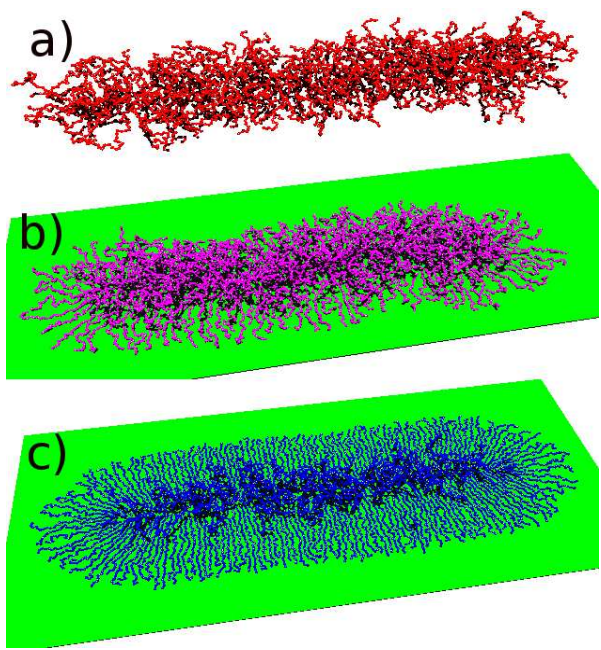}
\caption{Snapshots of equilibrated bottle-brushes with $L=121$ backbone monomers
and $244$ side chains of length $N=30$ ($\sigma_g=1$) displayed for different
adsorption strengths: (a) $\epsilon_s/k_BT$=0, (b) $\epsilon_s/k_BT= 4.0$, and
(c) $\epsilon_s/k_BT=8.0$. The spreading of the BB-molecule on the surface with
growing adhesion, $\epsilon_s/k_B > 0$ is clearly seen.}
\label{fig_meantension2}
\end{figure}

The substrate in the present investigation is considered simply as a
structureless adsorbing plane, with a Lennard-Jones potential acting with
strength $\epsilon_s$ in the perpendicular $z$--direction, $V^{\mbox{\tiny
LJ}}(z)=4\epsilon_s[(\sigma/z)^{12} - (\sigma/z)^6]$. In our simulations we
consider as a rule the case of {\it strong} adsorption, $\epsilon_s/k_BT=4.0
\div 8.0$. With these interaction between the monomers, and at different
degree of adhesion to the surface, we observe well equilibrated bottle-brush
molecules, as shown in Fig.~\ref{fig_meantension2}.

In reality the thermal energy is $k_BT\approx 4.11\times10^{-21}$~J/K
at absolute temperature $T\approx 300$~K, so for a typical Kuhn length of
$\sigma=1$~nm one obtains for the tension $f\approx 4$~pN. Estimates show
\cite{panyukovmacro} that this tension is too small to significantly change the
lifetime  $\tau\approx 10^{30}$ of covalent bonds with energy $\approx
100k_BT$. Adsorbed bottle-brush molecules on graphite or water/alcohol mixtures
with spreading coefficient $S\approx 20$~mN/m are capable of generating a force
of approximately $220$~pN \cite{sheiko} which reduces the lifetime to $10^4$~s.
In our coarse-grained model we consider bonds with dissociation energy of the
order of $10k_BT$. With Kuhn lengths $\approx 1$~nm and strength of adsorption
$\epsilon_s/k_BT \approx 1\div 8$ we find $\tau$ to be $10^2\div 10^3$ time
units (t.u.) whereby 1 t.u. is $\approx 10^{-12}$~s.

The bottle-brush dynamics is obtain by solving a Langevin equation for the
position $\mathbf q_n=[x_n,y_n,z_n]$ of each bead in the molecule,
\begin{equation}
m\ddot{\mathbf q}_n = \mathbf F_n^{j} + \mathbf F_n^{\mbox{\tiny
WCA}} -\gamma\dot{\mathbf q}_n + \mathbf R_n(t), \qquad (n,\ldots,N)
\label{langevin}
\end{equation}
which describes the Brownian motion of a set of bonded particles. In
Eq.~(\ref{langevin}) we choose the force $\mathbf F_n^{j}$ with $j=M$, or $KG$,
depending on the character of bonded (breakable -- Eq.~\ref{morse} or
nonbreakable -- Eq.~\ref{fene}) interaction.
The influence of solvent is split into slowly evolving viscous force and rapidly
fluctuating stochastic force. The random, Gaussian force $\mathbf R_n$ is
related to friction coefficient $\gamma$ by the fluctuation-dissipation theorem,
$\gamma=0.25$. The integration step is $0.002$ t.u. Time is
measured in units of $\sqrt{m\sigma^2/D}$ or $\sqrt{m\sigma^2/\epsilon}$
(where
$m$ denotes the mass of the
beads, $m=1$), depending on whether $V^{KG}(r)$ or $V^{M}$ is used.  We
emphasize at this point that in our coarse-grained model no
explicit solvent particles are included.
Our simulation was performed  in
the weakly damped regime of $\gamma=0.25$
as we modeled the interface between solid substrate and air. However, we carried
out also simulation in the strongly damped regime for $\gamma=10$. No
qualitative changes were discovered except an absolute overall increase of the
rupture times which is natural for a more viscous environment.

The initially created configurations, Fig.~\ref{flatbb} are equilibrated by MD
for a sufficiently long period of time so that the mean squared displacement of
the polymer center-of-mass moves a distance several $(3\div 5)$ times larger
than the polymer size. We then start the simulation with a well equilibrated
conformation - Fig.~\ref{centipede} - and measure the tension $f$ generated by
the steric repulsion of side chains on the covalent bonds that comprise the
macromolecule backbone.
Tension was measured by sampling the length of each bond and using derivative
of Eq.~(\ref{fene}).
In separate runs we allow
thermal scission of bonds by using the second bonded potential given by
Eq.~\ref{morse}. In that case we sample the frequency of bond breaking with
respect to the bond position in the backbone and create a Probability
Distribution  (a rupture probability histogram).

\section{MD results}

An important aspect of the tension-induced scission of bonds in a BB-molecule,
\begin{figure}[htb]
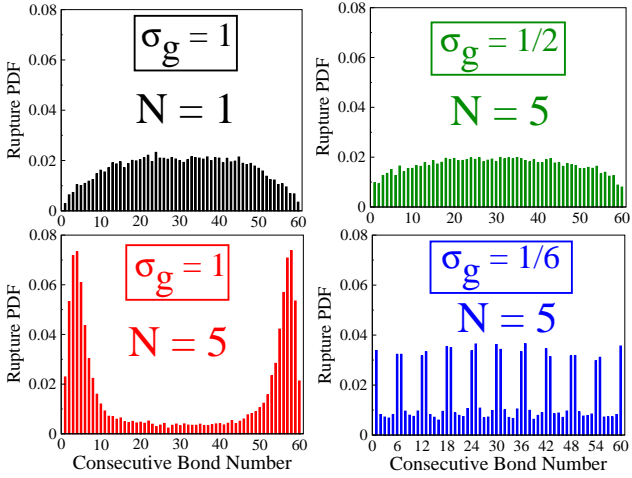

\onefigure[scale=0.31]{rupturePDF.eps}
\caption{Rupture probability histograms for a polymer backbone with $L=61$
presented for different values of grafting density $\sigma_g$ and length of side
chains $N$ as indicated. Here $\epsilon_s/k_BT=4.0$ and the bonded interaction
is governed by Morse potential given by Eq.~(\ref{morse}).}
\label{fig_rupturePDF}
\end{figure}
adsorbed on a solid plane is the probability distribution of rupture
(scission frequency) along the molecule backbone since it determines the
fragmentation kinetics which can be experimentally analyzed. In order to
interpret experimental results~\cite{lebedeva,park2} by means of scaling
theory~\cite{244,sheiko}, one assumes that tension of the side chains is
focused onto the backbone at its ends and transmitted through the
\begin{figure}[htb]
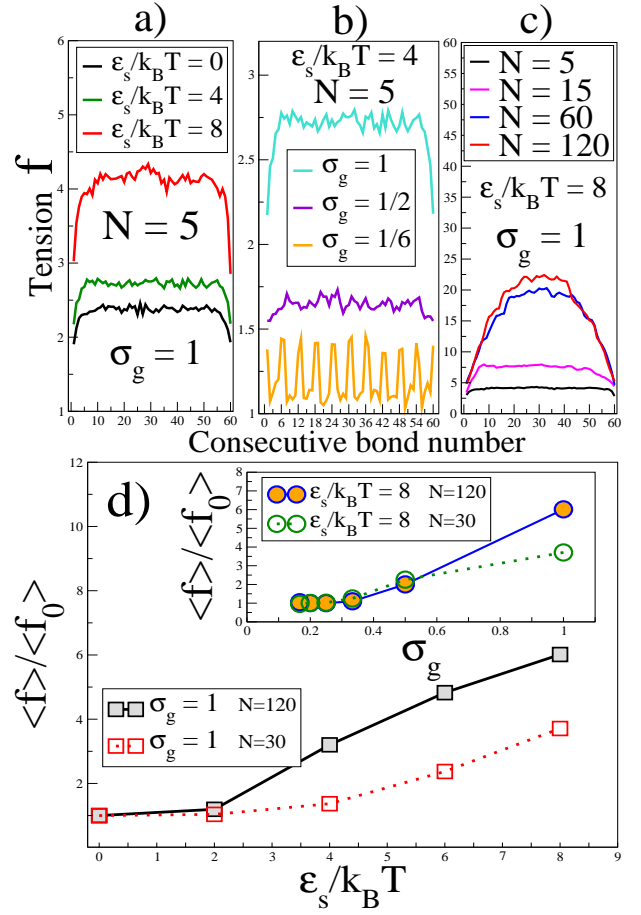

\onefigure[height=6.cm,width=8.cm]{tension_inc_L61.eps}
\onefigure[height=6.cm,width=8.cm]{tensionVSeps_L121.eps}
\caption{ The tension profile along the brush backbones made of $L=61$ particles
displayed as indicated for different: (a) adsorption strengths
$\epsilon_s/k_BT$,
(b) grafting densities $\sigma_g$, (c) lengths of side chains $N$. (d)
Variation of the mean tension $\langle
f\rangle/\langle f_0\rangle$ (in dimensionless units) generated in the backbone
of bottle-brush with $L=121$ as a function of adsorption strength
$\epsilon_s/k_{B}T$ and grafting density $\sigma_g$ (inset).}
\label{fig_f-profile}
\end{figure}
middle part of the backbone. This would then lead to uniform scission
probability of each bonds along the backbone. In Fig.~\ref{fig_rupturePDF} we
demonstrate that the rupture frequency depends, in fact, essentially on the
side chain length, $N$, provided grafting density $\sigma_g$ is held constant
(in an earlier paper we found that this frequency does not depend on backbone
length $L$ for sufficiently long BB-molecules~\cite{milchev}).
While the rupture frequency levels off and stays uniform in the middle for
$\sigma_g = 1$ and $N=1$, cf. upper left panel in Fig.~\ref{fig_rupturePDF},
one observes sharp focusing at both ends of the chain for $N=5$.
As argued in Ref.~\cite{milchev}, this effect is due to mutual
blocking of the side chains which are in the middle of bottle-brush molecule. In
contrast, side chains that are closer to the ends of the backbone are much more
mobile and, therefore, pull on the backbone more frequently. Thus, it is not
only the tension value but also the rate at which this tension is applied that
determine eventually the distribution of scission rates along the backbone.
However, for
$N=5$ and $\sigma_g = 1/2$ the distribution changes qualitatively, and attains
a shape as for $\sigma_g = 1.0,\;N=1$ again. Moreover, for smaller density of
grafting, $\sigma_g = 1/6$, one finds ``comb-like`` distribution which suggests
that bonds adjacent to the tethered backbone monomers break  then five
times more frequently.

\begin{figure}
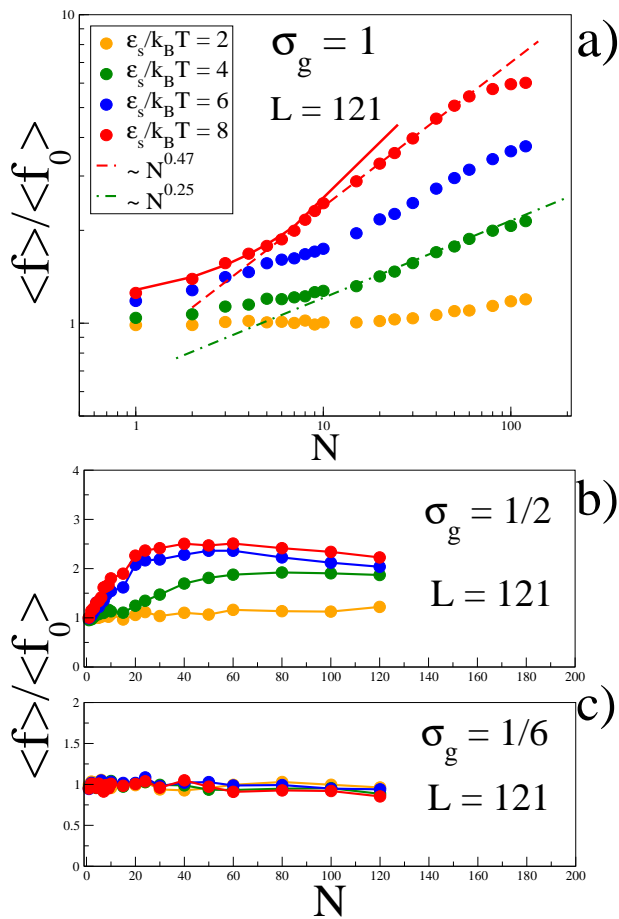

\onefigure[height=6.cm,width=8.cm]{f_N.eps}
\onefigure[height=6.cm,width=8.cm]{meantVSN_L121_gd26_pompon.eps}
\caption{(a)--(c) Mean (dimensionless) tension $\langle f\rangle$ in a
bottle-brush backbone composed of $L=121$ particles is plotted as a function of
the number of beads in the side chains $N$ for different adsorption strength
$\epsilon_s/k_{B}T$ (as indicated),  and grafting density: (a) $\sigma_g=1$, (b)
$\sigma_g=1/2$, (c) $\sigma_g=1/6$. Dashed lines represent linear fits with
different slope. 
}
\label{fig_meantension}
\end{figure}
The distribution of tension $f$ for different values of the governing
parameters $\epsilon_s, \;\sigma_g$ and $N$, is shown in
Fig.~\ref{fig_f-profile}a,~b,~c. Evidently, $f$ reaches a plateau away from
chain ends. From Fig.~\ref{fig_f-profile}d, the mean value $\langle
f\rangle$ is found to grow steadily with adsorption strength,
$\epsilon_s$ (for $\sigma_g=1$), and also with grafting density,
$\sigma_g$ (for $\epsilon_s/k_BT=8$).
Our simulation results yield a Gaussian distribution of $f$ around its mean
value $\langle f\rangle$ (not shown here).
 The tension is seen to focus on the middle of the chain with
increasing side chain length - Fig.~\ref{fig_f-profile}c. One should note again
the comb-like variation of $f$ over the successive bonds at $\sigma_g = 1/6$,
i.e. exactly where the side chains are grafted.

In Fig.~\ref{fig_meantension} we show one of the main results of this study -
the variation of tension in the bonds along the bottle-brush backbone with
growing length $N$ of the side chains whereby both the adsorption strength,
$\epsilon_s$ as well as the grafting density $\sigma_g$ are
varied. In all cases
$f$ is displayed in dimensionless units, taken as the ratio to mean bond
tension of a non-adsorbed BB-molecule $f_0$. Evidently, the theoretical
predictions~\cite{244} for a linear scaling relationship, $f \propto N$, appears
only for rather short side chains, $N < 10$ at  the largest density $\sigma_g =
1$ and adsorption strength $\epsilon_s = 8.0$ while for weaker adhesion,
$\epsilon_s = 4.0$, one finds $f \propto N^{1/4}$, as predicted~\cite{244}, yet
only for $10 \le N \le 60$.  For loosely grafted side chains, $\sigma_g \le 0.5$
and weaker adsorption,  $\epsilon_s = 2.0$, the tension becomes progressively
insensitive to $N$, saturating at a constant value at shorter and shorter length
of the side chains. At $\sigma_g=1/6$ the tension $f$ remains unaffected by
adsorption whatsoever for {\em any} $N$.

Generally, one observes progressive deviations from a power-law relationship for
longer side chains $N > 70$ even at the strongest adhesion. We believe that this
is due to a progressive immobilization of the BB-molecule for long side chains
which spread nearly parallel to one another, cf. Fig.~\ref{fig_meantension2},
and mutually block each others mobility at large values of $\epsilon_s$,
effectively ''freezing`` the macromolecule.

\begin{figure}
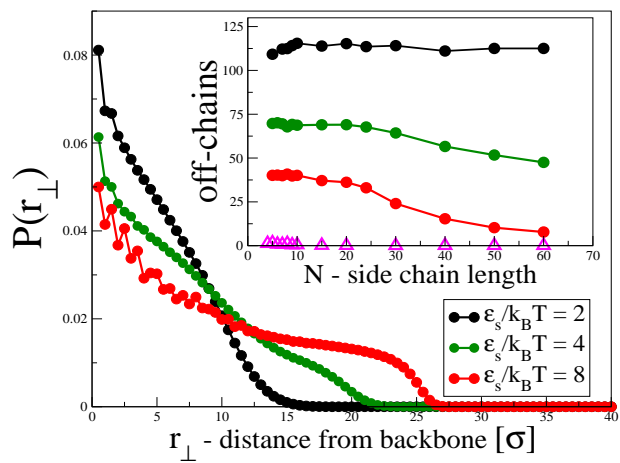

\onefigure[height=6.cm,width=8.cm]{density.eps}
\caption{ Monomer density profiles measured perpendicular to the backbone of
bottle-brushes with $L=121$  and $N=30$  displayed for different adsorption
strengths as
indicated in the legend. The inset presents average number of chains above the
substrate for a brush with $L=61$ as a function side chain length $N$ for
different sets of $\epsilon_s/k_BT$ and $\sigma_g$ (triangles stand for
$\epsilon_s/k_BT$=1 and $\sigma_g$=1/2) }
\label{fig_profiles}
\end{figure}

Finally, we show the density profile of all BB-monomers in direction
perpendicular to the backbone for several strengths of the adsorption potential
- Fig.~\ref{fig_profiles}. The shape of such cross-section profiles is important
because it has been measured experimentally by AFM for poly($n-$butyl acrylate)
side chains on mica (for strong adsorption and $\sigma_g=1$), and has also been
reproduced by Self-Consistent Field theory~\cite{244}.
A close inspection of the cross-section profiles, shown in
Fig.~\ref{fig_profiles}, reveals qualitatively the same shapes as predicted
theoretically~\cite{244}: (i) for weak adsorption, $\epsilon_s/k_BT=2.0$, one
observes a nearly triangular ''tent-like`` profile; (ii) for moderate adhesion,
$\epsilon_s/k_BT=4.0$, a shoulder of monolayer height is seen to form while most
of the monomers still stay in the cap, and (iii), at the strongest attraction,
$\epsilon_s/k_BT=8.0$, one finds a well-expressed monolayer and a narrow sharp
cap (a cusp). Instead of a cusp right above the molecule backbone, however, the
AFM experiment yields a well-defined parabolic maximum~\cite{244}. This is most
probably due to the deformation of the BB-cap by the AFM cantilever (about 20\%
of its height~\cite{244}) which thus modifies and distorts its true morphology.
In fact, the sharp cusp-like cap is formed by partitioning of side chains into
adsorbed and non-adsorbed ones at high grafting density -
Fig.~\ref{fig_meantension2}b,~c. Indeed, some of the side chains may relieve the
deformation in the strongly adsorbed monolayer and minimize the BB free energy
by lifting off the substrate. In the inset to Fig.~\ref{fig_profiles} one may
see the variation in the number of such off-chains with length $N$ at different
degrees of adhesion. For strong adsorption, $\epsilon_s/k_BT=8.0$, the number of
off-chains decreases with $N$ since the energetic cost of desorption becomes too
high. One can also see  that for $\sigma_g = 1/2,\; \epsilon_s=1.0$ the number
of off-chains  is close to zero since no crowding of monomers on the surface
takes place. It should be clear that these 'dangling' off-chains bring about the
sharp cap in the cross-section density profile, observed in
Fig.~\ref{fig_profiles}.

\section{Concluding remarks}

In this Letter we report on simulation results from our study of bottle-brush
macromolecules adsorbed on a structureless attractive surface and examine
conditions which govern the build-up of bond tension and the ensuing frequency
of bond scission distribution along the molecule backbone. Results are
compared to those derived from experiments and theoretical considerations.  Our
findings can be summarized as follows:
\begin{itemize}
\item The variation of bond tension $f$ with growing length of the side chains
$N$, adhesion strength $\epsilon_s$, and grafting density $\sigma_g$ comply
with some important theoretical predictions~\cite{244} albeit in a rather
narrow interval. Especially, a scaling law $f \propto f_{\mbox{\tiny S}} N$ is
observed for
strong adsorption and very short side chains $N < 10$ only. At moderate
adsorption one also finds $f \propto f_{\mbox{\tiny S}}N^{1/4}$ in agreement
with
theory~\cite{244}, yet for $N < 70$.  Further increase of $N$ slows down the
tension build-up and eventually makes $f$ insensitive to $N$ as the side chains
mutually block each other and the brush molecule gets progressively ''frozen``.

\item The distribution of tension and the scission frequency along the bonds of
the bottle-brush backbone change strongly with $N$ and $\sigma_g$. Different
combinations of these parameters can induce uniform, comb-like, or strongly
focused tension distribution in adsorbed BB polymers.

\item The cross-section monomer density profile of adsorbed bottle-brush
molecules agrees with experimental observations and theoretical predictions of
the shape at different regimes of adsorption. We find, however, the existence
of sharp cusp-like peak in the profiles which manifests side chain partitioning
into adsorbed and off-surface chains at strong adhesions and large grafting
density.

\end{itemize}

\acknowledgments
We thank V.G.~Rostiashvili for fruitful discussions. A.M.~gratefully
acknowledges support by the Max-Planck-Institute for Polymer Research in Mainz
during the time of this investigation. This research has been supported by the
Deutsche Forschungsgemeinschaft (DFG), Grants SFB 625/B4 and FOR 597.

\end{document}